\documentclass[letterpaper, 10pt, conference, twocolumn]{IEEEtran}
\usepackage{fancyhdr}
\usepackage{amsmath,epsfig}
\usepackage{threeparttable}
\usepackage{epsf,epsfig}
\usepackage{amsthm}
\usepackage{amsmath}
\usepackage{amssymb}
\usepackage{amsfonts}
\usepackage{algorithmic}
\usepackage{algorithm}
\usepackage[noadjust]{cite}
\usepackage{dsfont}
\usepackage{subfigure}

\newcommand{\TT}{\mathsf{T}}
\newcommand{\RR}{\mathsf{R}}

\newtheorem{lemma}{Lemma}

\newtheorem{proposition}{Proposition}

\def\E{\mathbb{E}}

\def\phi{\varphi}

\def\SINR{\mathsf{SINR}}

\def\l{\left}
\def\r{\right}
\def\({\left(}
\def\){\right)}

\def\b0{{\mathbf{0}}}

\def\bB{{\mathbf{B}}}
\def\bC{{\mathbf{C}}}

\def\bF{{\mathbf{F}}}
\def\bG{{\mathbf{G}}}
\def\bH{{\mathbf{H}}}

\def\bU{{\mathbf{U}}}
\def\bV{{\mathbf{V}}}

\def\bX{{\mathbf{X}}}

\newcommand{\Pout}{P_{\mathsf{out}}}

\newcommand{\diag}{\mathrm{diag}}

\newcommand{\nn}{\nonumber}

\begin{document}

\title{\huge \setlength{\baselineskip}{30pt} Cooperative Feedback  for  MIMO Interference Channels}
\author{
\authorblockN{Kaibin Huang}
\authorblockA{School of EEE, Yonsei University\\
Seoul, Korea\\
Email: huangkb@yonsei.ac.kr\vspace{-5pt}}
\and
\authorblockN{Rui Zhang}
\authorblockA{Institute for Infocomm Research, A*STAR, Singapore\\
ECE Department, National University of Singapore\\
Email:  rzhang@i2r.a-star.edu.sg}
} \maketitle

\begin{abstract} Multi-antenna precoding effectively  mitigates the interference in  wireless networks. However, the precoding efficiency can be significantly degraded by the overhead due to the required  feedback of channel state information (CSI). This paper addresses such an issue by proposing a systematic method of designing precoders for the two-user multiple-input-multiple-output (MIMO) interference channels based on finite-rate CSI feedback from receivers to their interferers, called \emph{cooperative feedback}. Specifically, each precoder is decomposed into inner and outer precoders for nulling interference and improving the data link array gain, respectively. The inner precoders are further designed to suppress residual interference resulting from finite-rate cooperative feedback. To regulate residual interference due to precoder quantization, additional scalar cooperative feedback signals are designed to control transmitters' power using different criteria including applying interference margins, maximizing sum throughput,  and minimizing outage probability. Simulation shows that such additional  feedback effectively alleviates performance degradation due to quantized precoder feedback. 
\end{abstract}

\section{introduction}
In multi-antenna wireless networks, precoding can effectively mitigate  interference between coexisting links. This paper presents a new approach of efficiently implementing precoding   in the two-user  multiple-input-multiple-output (MIMO) interference channels  by exchanging finite-rate channel state information (CSI). Specifically, precoders are designed to suppress interference to the interfered receivers based on their quantized CSI feedback, and the residual interference is regulated by additional feedback of  power control signals. 

Recently,  progresses have been made in analyzing  the capacity   of multi-antenna  interference channels. In particular,  interference alignment techniques have been proposed for achieving the channel capacity for high signal-to-noise ratios (SNRs)   \cite{CadJafar:InterfAlignment:2007}. Such techniques, however, have limited practicality due to their complexity, requirement of perfect global CSI and their sub-optimality for finite SNRs. This prompts the development of linear precoding algorithms for practical decentralized wireless networks \cite{ZakhourGesbert:DistMutlicellMISOPrecodingLayerVirtualSINR, ChaeHeath:InterfAwareCoordBeamformTwoCell, ZhangCui:CoopIMMISOBeamform, DahYu:CoordBeamformMulticell:2010}. For time-division multiplexing (TDD)  multiple-input-single-output (MISO) interference channels, it is proposed in \cite{ZakhourGesbert:DistMutlicellMISOPrecodingLayerVirtualSINR, DahYu:CoordBeamformMulticell:2010} 
that   forward-link beamformers can be adapted distributively based on reverse-link  signal-to-interference-pluse-noise ratios (SINRs).   
Targeting the two-user MIMO interference channels, linear transceivers are designed  in \cite{ChaeHeath:InterfAwareCoordBeamformTwoCell} under the constraint of one data stream per user and using different criteria including zero-forcing and minimum mean square error. In \cite{ZhangCui:CoopIMMISOBeamform},  the rate region for MISO interference channels is analyzed based on the cognitive radio principle, yielding  
a  message passing  algorithm for enabling distributive  beamforming. The above prior work does not address the issue of finite-rate feedback though it is widely used in the practice to enable precoding. Neglecting  feedback CSI errors in precoder designs 
 leads to over optimistic network performance. 

For MIMO precoding  systems, the substantiality of  CSI feedback overhead has motivated extensive   research on CSI quantization  algorithms, forming a field called \emph{limited feedback} \cite{Love:OverviewLimitFbWirelssComm:2008}. Recent limited feedback research has focused on  MIMO downlink systems, where multiuser CSI feedback  supports \emph{space division multiple access} (SDMA) \cite{Gesbert:ShiftMIMOParadigm:2007}.  It has been found that the number  of feedback bits  per user has to increase with the transmit SNR so as to bound the throughput loss caused by feedback quantization \cite{Jindal:MIMOBroadcastFiniteRateFeedback:06}.
Furthermore, such a loss  can be reduced  by exploiting \emph{multiuser diversity} \cite{SharifHassibi:CapMIMOBroadcastPartSideInfo:Feb:05, Huang:OrthBeamSDMALimtFb:07}. Designing limited feedback algorithms  for interference channels is more challenging  due to the decentralized network architecture and the growth of feedback CSI. Cooperative feedback algorithms are proposed in \cite{HuangZhang:CoopFeedbackCognitiveRadio} for 
a two-user  cognitive radio network, where the secondary   transmitter adjusts its beamformer to suppress interference to  the primary   receiver that cooperates by feedback to the secondary transmitter.  This design is tailored for a MISO cognitive radio network and thus unsuitable for general MIMO interference channels, which motivates the current work.

We  consider two coexisting MIMO links where all nodes employ equal numbers of antennas and  linear precoding is enabled by quantized cooperative feedback.  Channels are assumed to have i.i.d. Rayleigh fading. 
A systematic method is proposed for jointly designing linear precoders and equalizers  under an orthogonality constraint, which  decouples the links in the case of perfect  feedback. To be specific, precoders and equalizers are decomposed into inner and outer components,  where the former are designed to suppress residual interference caused by feedback errors and the latter  to   enhance link array gain. 
Second, additional  scalar cooperative feedback, called \emph{interference power control} (IPC) feedback, is proposed for controlling  transmitters'  power so as to regulate the residual interference. Specifically, the IPC feedback algorithms are designed using  different criteria  including fixed interference margin, maximum sum throughput,  and minimum  outage probability.

{\bf Notation:} The superscript $\dagger$ represents matrix Hermitian  transpose. The operator $[\bX]_{k}$ gives the $k$th column of a matrix $\bX$. Let $\preceq$, $\prec$,  $\succeq$ and $\succ$ represent element-wise inequalities between two real vectors. 

\section{System Model}\label{Section:System}
\begin{figure}[t]
\begin{center}
\includegraphics[width=9cm]{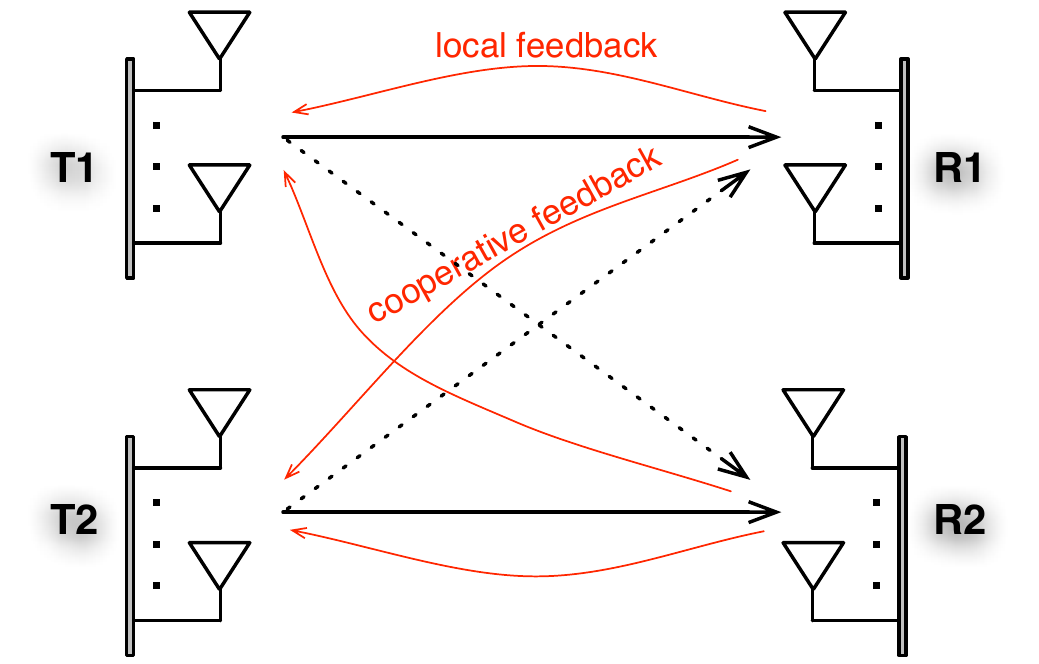}
\caption{MIMO interference channels with local and cooperative feedback.   }
\label{Fig:Sys}
\end{center}
\end{figure}

 We consider two interfering wireless links as illustrated in Fig.~\ref{Fig:Sys}, where the two pairs of transceivers are denoted as $(\TT_1, \RR_1)$ and $(\TT_2, \RR_2)$. Each transmitter/receiver has  $L$  antennas employed for suppressing the interference  as well as supporting spatial multiplexing. These functions require CSI feedback from receivers to their interferers and intended transmitters, called \emph{cooperative feedback} and \emph{local feedback}, respectively. 
 We assume perfect   CSI estimation and  local feedback, allowing the current design to focus on suppressing interference 
 caused by  cooperative feedback quantization. All channels are assumed to follow independent blocking fading.  The channel coefficients are samples of  i.i.d. $\mathcal{CN}(0,1)$ random variables.   Let $\bH_{mn}$ denote a $L\times L$ i.i.d. $\mathcal{CN}(0,1)$ matrix representing  fading of the channel from $\TT_n$ to $\RR_m$. Then the interference channels are modeled as  $\{\nu\bH_{mn}\}$ and the data channels as  $\{\bH_{mm}\}$ where $m,n\in\{1,2\}$ and $m\neq n$. The factor $\nu < 1$ quantifies the difference in transmission distance between the data and interference links. 

Each link supports  $M\leq L$ spatial data streams by linear precoding and equalization. To regulate residual interference caused by precoder feedback errors, the total transmission power of each transmitter is   constrained by cooperative IPC feedback. For simplicity, the scalar IPC feedback is assumed perfect since it requires much less overhead than the precoder feedback.    Each transmitter uses identical transmission  power for  all spatial streams,  represented by  $P_n$  for   $\TT_n$ with $n=1, 2$ and the maximum $P_{\max}$. Assume that all  additive white noise samples are i.i.d.  $\mathcal{CN}(0, 1)$ random variables. Let $\bG_m$ and $\bF_m$ denote the linear equalizer used by $\RR_m$ and the linear precoder applied at $\TT_m$, respectively.  
Thus the  receive  signal-to-interference-plus-noise ratio (SINR)   at  $\RR_m$ for the $\ell$th stream can be written as 
\begin{equation}
\SINR_m^{[\ell]}  := \frac{ P_m|[\bG_m]_\ell^\dagger \bH_{mm}[\bF_m]_\ell |^2}{1 + P_n\nu \|[\bG_m]_\ell^\dagger \bH_{mn}\bF_n\|^2 }, \quad m\neq n. 
\label{Eq:SINR}
\end{equation}

Two performance metrics, ergodic throughput  and outage probability, are considered. The total ergodic throughput of both links, called \emph{sum throughput}, is defined as 
\begin{equation} \label{Eq:ErgCap}
\bar{C} := \sum_{m=1}^2\sum_{\ell =1}^M \E\l[\log_2\l(1+\SINR_m^{[\ell]} \r)\r] 
\end{equation}
where $\SINR_m^{[\ell]}$ is  given in \eqref{Eq:SINR}. 
Next, consider the scenario where the coding rates for all data streams are fixed  at $\log_2(1+\theta)$ where $\theta$ is the receive SINR threshold for correct decoding. We define an outage event as one that the SINR of at least one data stream is smaller than $\theta$. It follows that the  outage probability  is given by
\begin{equation}
\Pout :=\Pr\l(\min_{m=1,2}\min_{1\leq \ell\leq M}\SINR_m^{[\ell]}\leq \theta\r).  \label{Eq:Pout}
\end{equation}

\section{Precoding with Limited Feedback}\label{Section:Orth:Algo}

\subsection{Precoder Design}
A  pair of precoder and equalizer $(\bG_m, \bF_n)$ with $m\neq n$ are jointly designed under the following orthogonality constraint 
\begin{equation}\label{Eq:OrthConst}
\bG_m^\dagger \bH_{mn} \bF_n = \mathbf{0}, \quad \ m,n\in\{1, 2\}, m \neq n. 
\end{equation}
The constraint aims at decoupling  the links and requires that $L \geq 2M$.  A key step of  the proposed design is to decompose the precoder $\bF_n$ into  an \emph{inner precoder} $\bF_n^{\mathsf{i}}$ and an \emph{outer precoder} $\bF_n^{\mathsf{o}}$. Specifically, $\bF_n  = \bF_n^{\mathsf{i}}\bF_n^{\mathsf{o}}$ where  $\bF^{\mathsf{i}}_n$  and $\bF_n^{\mathsf{o}}$ are $L\times M$ and $M\times M$ matrices, respectively, where the size of  $\bF^{\mathsf{i}}_n$ is minimized to reduce feedback overhead. Similarly, we decompose the equalizer $\bG_m$ as $\bG_m  = \bG_m^{\mathsf{i}}\bG_m^{\mathsf{o}}$ where  $\bG_m^{\mathsf{i}}$ is a $L\times N$ \emph{inner equalizer} and  $\bG_m^{\mathsf{o}}$  a $N \times M$ \emph{outer equalizer}, where $N$ is a design parameter under the constraints  $N\geq M$ and $N \leq L-M$.  
The  inner precoder/equalizer   pair  $(\bG_m^{\mathsf{i}}, \bF^{\mathsf{i}}_n)$ is designed to  enforce the constraint in \eqref{Eq:OrthConst} while the outer pair  $(\bG_m^{\mathsf{o}}, \bF^{\mathsf{o}}_n)$ enhances the link array gain as discussed in the sequel. It follows that 
\begin{equation}\label{Eq:OrthConst:b}
(\bG_m^{\mathsf{i}})^\dagger \bH_{mn} \bF_n^{\mathsf{i}} = \mathbf{0},\quad  m\neq n. 
\end{equation}
Under this constraint, $(\bG_m^{\mathsf{i}}, \bF^{\mathsf{i}}_n)$ are  designed by decomposing $\bH_{mn}$ using the 
singular value decomposition (SVD) as 
\begin{equation}
\bH_{mn} = \bV_{mn} \l[ \begin{matrix}   \sqrt{\lambda_{mn}^{[1]}} & &\mathbf{0}  \\   
& \ddots& \\
\mathbf{0}& &\sqrt{\lambda_{mn}^{[L]}}\end{matrix} \r] \bU_{mn}^\dagger, \quad m\neq n   \label{Eq:H:SVD:a}
\end{equation}
where the unitary matrices  $\bV_{mn}$ and $\bU_{mn}$ consist of the left and right singular vectors of $\bH_{mn}$ as columns, respectively, and $\l\{\lambda_{mn}^{[\ell]}\r\}$ denote  the  eigenvalues of $\bH_{mn}\bH_{mn}^\dagger$ following the descending order. Let $\mathcal{A}$ and $\mathcal{B}$ be two subsets of the indices $\{1, 2, \cdots, L\}$ with $|\mathcal{A}|=N$, $|\mathcal{B}| = M$,  and $\mathcal{A}\cap \mathcal{B} = \emptyset$.  
  The constraint in \eqref{Eq:OrthConst:b} can be satisfied by  choosing  
\begin{equation}
\bG_m^{\mathsf{i}} = \{[\bV_{mn}]_k\mid k\in \mathcal{A}\}\ \textrm{and}\ \bF_n^{\mathsf{i}}=\{[\bU_{mn}]_k\mid k\in \mathcal{B}\}. \label{Eq:Inner:Design:a}
\end{equation}

Given $(\bG_m^{\mathsf{i}}, \bF_m^{\mathsf{i}})$, the outer pair $(\bG_m^{\mathsf{o}}, \bF_m^{\mathsf{o}})$ are jointly designed based on the SVD of the $N\times M$  effective channels $\bH_{mm}^{\mathsf{o}} := \bG_m^{\mathsf{i}}\bH_{mm}\bF_m^{\mathsf{i}}$:
\begin{equation}
\bH_{mm}^{\mathsf{o}} = \bV_{mm} \l[ \begin{matrix}   \sqrt{\lambda_{mm}^{[1]}} & &\mathbf{0}  \\   
& \ddots& \\
\mathbf{0}& &\sqrt{\lambda_{mm}^{[M]}}\\
\mathbf{0}&\cdots&\mathbf{0}\end{matrix} \r] \bU_{mm}^\dagger.   \label{Eq:H:SVD:11}
\end{equation}
 Note that the elements of $\bH_{mm}^{\mathsf{o}}$ are i.i.d. $\mathcal{CN}(0, 1)$ random variables and their distributions are independent of $(\bG_m^{\mathsf{i}}, \bF_m^{\mathsf{i}})$ since $\bH_{mm}$ is isotropic.  To transmit data through the eigenmodes of $\bH_{mm}$,  $\bG_m^{\mathsf{o}}$ and  $\bF_m^{\mathsf{o}}$ are chosen as 
\begin{equation}
\bG_m^{\mathsf{o}} = \l\{[\bV_{mm}]_1, [\bV_{mm}]_2, \cdots, [\bV_{mm}]_M\r\}\ \textsf{and}\ \bF_m^{\mathsf{o}} = \bU_{mm}. \nn
\end{equation} 
With perfect CSI feedback, the above  precoder and equalizer joint design converts each data link into $M$ parallel spatial channels which are free of interference.

Note that increasing $N$ enhances the array gain of both links. Specifically, the expectations of the SNRs increase with $N$.  Thus, $N$ should take its maximum  $(L-M)$. However, maximizing $N$ need not be optimal for the link performance in the case of quantized feedback as discussed in the sequel.

\subsection{Quantized Precoder Feedback} 
In this section, we choose the index sets $\mathcal{A}$ and $\mathcal{B}$  in \eqref{Eq:Inner:Design:a} with the objective of suppressing the residual interference caused by precoder feedback errors. 

Recall that the precoding at  $\TT_n$ is  enabled by quantized cooperative feedback of $\bF_n^{\mathsf{i}}$ from $\RR_m$ with $m\neq n$. Let $\hat{\bF}_n^{\mathsf{i}}$ denote the quantized version of  $\bF_n^{\mathsf{i}}$ and define the  resultant quantization error $\epsilon_n$  as 
 \begin{equation}\label{Eq:QuantErr}
\epsilon_n := 1 - \frac{\|(\bF_n^{\mathsf{i}})^\dagger \hat{\bF}_n^{\mathsf{i}}\|^2_{\mathsf{F}}}{M}, \quad n = 1, 2
\end{equation}
where  $0\leq  \epsilon_n \leq 1$. The  error $\epsilon_n$ is zero in the case of perfect cooperative feedback, namely $\bF_n^{\mathsf{i}} = \hat{\bF}_n^{\mathsf{i}}$. A nonzero error results in  violation of the orthogonality constraint in \eqref{Eq:OrthConst:b}
\begin{equation}\label{Eq:OrthConst:a}
\bG_m^{\mathsf{i}}\bH_{mn}\hat{\bF}_n^{\mathsf{i}} \neq 0,\quad m\neq n. 
\end{equation}
The resultant   residual interference  from $\TT_n$ to the $\ell$th data stream of  $\RR_m$ has the power 
\begin{equation}
I_{mn}^{[\ell]} := P_n\nu\| [\bG_m^{\mathsf{o}}]_{\ell}^\dagger (\bG_m^{\mathsf{i}})^\dagger \bH_{mn} \hat{\bF}_n^{\mathsf{i}}\bF_n^{\mathsf{o}}\|^2,\quad  m \neq n.  
\label{Eq:Interf}
\end{equation}

Next, we choose  $\mathcal{A}$ and $\mathcal{B}$ in \eqref{Eq:Inner:Design:a} by minimizing an upper bound on  the residual interference power as follows. 
Based on \eqref{Eq:Inner:Design:a}, \eqref{Eq:H:SVD:a} can be rewritten as 
\begin{equation}
\bH_{mn} = \l[\begin{matrix}\bB_m&\bG^{\mathsf{i}}_{m}\end{matrix}\r]  \boldsymbol{\Sigma}_{mn}\l[\begin{matrix}  \bF^{\mathsf{i}}_n & \bC_n \end{matrix} \r]^\dagger  \label{Eq:H:SVD}
\end{equation}
where 
$
 \boldsymbol{\Sigma}_{mn}:= \boldsymbol{\Pi}\ \diag\l(\sqrt{\lambda_{mn}^{[1]}}, \sqrt{\lambda_{mn}^{[1]}}, \cdots,  \sqrt{\lambda_{mn}^{[L]}}\r) \boldsymbol{\Pi}^\dagger\label{Eq:Diag:Matrix}
$
with  $\boldsymbol{\Pi}$ being  an arbitrary  permutation matrix that rearranges the order of the singular values along the matrix diagonal. The columns of the matrices $\bB_m$ and $\bC_n$ comprise  the $(L-N)$ left and $(L-M)$ right singular vectors of $\bH_{mn}$, respectively, which are determined by $\boldsymbol{\Pi}$. 
Let the set $\mathcal{D}_{mn}$ contain the last $N$ elements  along the diagonal of $\boldsymbol{\Sigma}_{mn}$.
\begin{lemma} \label{Lem:Interf:Ub} The interference power $I_{mn}^{[\ell]}$ in \eqref{Eq:Interf} can be upper bounded as 
\begin{equation}
I_{mn}^{[\ell]}\leq M\nu P_n\epsilon_n\max_{\alpha\in\mathcal{D}_{mn} }\alpha^2,\quad m \neq n. \label{Eq:Interf:Ub}
\end{equation}
\end{lemma}
Readers can refer to the full paper \cite{HuangZhang:CoopMIMOInterfChannelLimFb} for the proofs of  the above  lemma as well as analytical results in the sequel. Minimizing the upper bound in \eqref{Eq:Interf:Ub} gives that $\mathcal{D}_{mn}$ consists of  the $N$ smallest singular  values of $\bH_{mn}$. Equivalently, $\boldsymbol{\Pi}$  is an identity matrix and thus  $\bG_m^{\mathsf{i}}$ and $\bF_n^{\mathsf{i}}$  are given as 
\begin{equation}\begin{aligned}
&\bG_m^{\mathsf{i}} &=&\l\{[\bV_{mn}]_{L-N+1}, [\bV_{mn}]_{L-N+1}, \cdots, [\bV_{mn}]_L\r\}\\
&\bF_n^{\mathsf{i}} &=&\l\{[\bU_{mn}]_1, [\bU_{mn}]_2, \cdots, [\bU_{mn}]_M\r\}. 
\end{aligned}\label{Eq:Inner:Design}
\end{equation}
Then \eqref{Eq:Interf:Ub} can be simplified as 
\begin{eqnarray}
I_{mn}^{[\ell]} \leq  M\nu P_n\lambda_{mn}^{[L-N+1]}\epsilon_n, \quad \forall \ 1\leq \ell\leq M, m\neq n. \label{Eq:Interf:Ub:c}
\end{eqnarray}
Note that the above upper bound on $I_{mn}^{[\ell]}$ is independent of the stream index $\ell$. On one hand, the upper bound reduces with decreasing  $N$.  On the other hand, as  mentioned earlier, larger $N$ increases link array gain. These opposite  effects of $N$ on link performance make it an important parameter for precoder optimization.  Finding the optimal $N$ is mathematically intractable but a numerical search is straightforward.

\section{Interference Power Control  Feedback} \label{Section:IPC}

\subsection{Fixed Interference Margin}\label{Section:FixMargin}
The receiver $\RR_m$ sends the IPC signal, denoted as $\eta_n$, to the interferer $\TT_n$ for controlling its transmission power as 
\begin{equation}
P_n  = \min(\eta_n, P_{\max}), \quad n = 1, 2. \label{Eq:TxPwr:IPC}
\end{equation}
The scalar $\eta_n$ is  designed to prevent the per-stream interference power at $\RR_m$  from exceeding a fixed margin $\tau$ with $\tau > 0$, namely $I_{mn}^{[\ell]} \leq \tau$ for all $0\leq \ell \leq M$. A sufficient condition for satisfying such constraints is to upper bound   the right hand side of \eqref{Eq:Interf:Ub:c}  by   $\tau$. It follows that 
\begin{equation}
\eta_n := \frac{\tau}{M\nu\lambda_{mn}^{[L-N+1]}\epsilon_n}, \quad m\neq n. \label{Eq:IPC}
\end{equation}
Given $\tau$, a lower bound $A_{\mathsf{IM}}$ on the  sum throughput  $\bar{C}$, called the \emph{achievable throughput},   is  obtained from \eqref{Eq:ErgCap} as 
\begin{equation}
A_{\mathsf{IM}} =  \sum_{m=1}^2\sum_{\ell =1}^M \log_2\l(1 + \frac{\min(\eta_m, P_{\max}) \lambda_{mm}^{[\ell]}}{1+\tau}\r)  \label{Eq:SumCap:LB}
\end{equation}
where $\eta_m$ is given  \eqref{Eq:IPC}.

It is infeasible to derive  the optimal value of $\tau$ for either maximizing $A_{\mathsf{IM}}$ in \eqref{Eq:SumCap:LB} or minimizing $\Pout$ in  \eqref{Eq:Pout}. However, for $P_{\max}$ being either large or small, simple insight into choosing $\tau$ can be derived   as follows. The residual interference power decreases continuously with reducing $P_{\max}$. Intuitively, $\tau$ should be kept small for small $P_{\max}$. For large $P_{\max}$, the choice of $\tau$ is less intuitive  since large $\tau$ lifts the constraints on the transmission power  but  causes stronger interference and vice versa. We show below that large $\tau$ is preferred for large $P_{\max}$. 
Let $\acute{\lambda}_k$ denote the eigenvalue of the Wishart matrix $\bH\bH^\dagger$ with $\bH$ being an i.i.d. $N\times M$ $\mathcal{CN}(0, 1)$ matrix. Define $\check{\lambda}_k$ similarly but  with $\bH$ being a $L\times L$ matrix. 
\begin{lemma} \label{Lemma:SumCap:IM} For large $P_{\max}$,  the achievable throughput is  
\begin{equation}\nn
A_{\mathsf{IM}}  =
 2\sum_{\ell = 1}^M  \E\l[\log_2\l(1\!+ \!\frac{\tau\acute{\lambda}_\ell}{(1+\tau)M\nu\check{\lambda}_{L-N+1}\epsilon_1}\r)\r] + o\l(1\r).
\end{equation}
\end{lemma}
\noindent It can be observed from the above result that  the first order term of $A_{\mathsf{IM}}$ attains its maximum for $\tau\rightarrow\infty$. However, this term  is finite even for asymptotically large  $P_{\max}$ and $\tau$, which is the inherent effect of residual interference.  
\begin{lemma} \label{Lem:Pout:IM}For large $P_{\max}$,  the outage probability  is upper bounded as  
\begin{equation}
\Pout  \leq2\Pr\l( \frac{\tau}{1+\tau}\times \frac{\acute{\lambda}_{\ell}}{M\nu\check\lambda_{L-N+1}\epsilon_1}< \theta \r) + o(1).\nn  \label{Eq:PoutIM:LargeP}
\end{equation}
\end{lemma}
\noindent Similar remarks on Lemma~\ref{Lemma:SumCap:IM} apply to Lemma~\ref{Lem:Pout:IM}.

\subsection{Sum Throughput  Criterion}\label{Section:IPC:Cap} \label{Section:ErgCap}
In this section, an iterative  IPC algorithm is designed for increasing the sum throughput  $\bar{C}$ in 
\eqref{Eq:ErgCap}. 
Since  $\bar{C}$ is a non-convex function of transmission power, directly maximizing $\bar{C}$ does not yield a simple IPC algorithm. Thus, we resort  to maximizing a lower bound $A_{\mathsf{ST}}$ (achievable throughput) on $\bar{C}$ instead,  obtained  from \eqref{Eq:ErgCap} and \eqref{Eq:Interf:Ub:c}  as 
$A_{\mathsf{ST}} =  \E\l[A\r]$ with 
\begin{equation}\label{Eq:Cap:Lb}\begin{aligned}
&A &:=& \ \sum_{\ell=1}^M\left[  \log_2\l(1+ \frac{P_1 \lambda_{11}^{[\ell]}}{1 + P_2 M\nu \lambda_{12}^{[L-N+1]} \epsilon_2}\r)+\right.\\
&&&\left. \log_2\l(1+\frac{ P_2 \lambda_{22}^{[\ell]}}{1 + P_1M\nu \lambda_{21}^{[L-N+1]}\epsilon_1}\r)\r]. 
\end{aligned}
\end{equation}
Thus, the optimal transmission power pair is given as 
\begin{equation}
(P_1^\star, P_2^\star) = \max_{P_1, P_2\in [0, P_{\max}]} A(P_1, P_2). 
\end{equation}
The objective function  $A$ remains non-convex and its maximum has no known closed-form. However, inspired by the message passing algorithm in \cite{ZhangCui:CoopIMMISOBeamform},   a  sub-optimal search for $(P_1^\star, P_2^\star)$  can be derived using  the fact that 
\[
\frac{\partial A(P_1^\star, P_2^\star)}{\partial P_m}   = 0 \ \forall\ m = 1, 2. 
\]
To this end, the slopes of $A$ are obtained using \eqref{Eq:Cap:Lb} as 
\begin{equation}
\frac{\partial A(P_1, P_2)}{\partial P_m}  = \mu_{m} + \psi_m - \rho_{m}\label{Eq:CapSlope}
\end{equation}
where 
\begin{eqnarray}
\mu_{m} &:=& \log_2e\sum\nolimits_{\ell=1}^M \frac{\lambda_{mm}^{[\ell]}}{1 + M\nu\lambda_{mn}^{[L-N+1]}\epsilon_nP_n + \lambda_{mm}^{[\ell]}P_m}\nn\\
\psi_{m} &:=& \log_2e\sum\nolimits_{\ell=1}^M \frac{M\nu\lambda_{nm}^{[L-N+1]}\epsilon_m}{1 + M\nu\lambda_{nm}^{[L-N+1]}\epsilon_mP_m + \lambda_{nn}^{[\ell]}P_n}\nn\\
\rho_{m} &:=&   \frac{\log_2e M^2\nu\lambda_{nm}^{[L-N+1]}\epsilon_m}{1+M\nu\lambda_{nm}^{[L-N+1]}\epsilon_mP_m}.   \nn
\end{eqnarray}
Note that based on available  CSI,  $\mu_{m}$ has to be  computed at $R_m$ and  $(\psi_{m}, \rho_m)$ at $R_n$ with $n\neq m$. Therefore, 
based on \eqref{Eq:CapSlope}, an iterative IPC feedback algorithm can be designed to have the following procedure. 

\noindent {\bf Algorithm $\mathbf{1}$:}
\begin{enumerate}
\item The transmitters $\TT_1$ and $\TT_2$ arbitrarily select the initial values for $P_1$ and $P_2$, respectively. 

\item The transmitters broadcast their choices of transmission power to the receivers. 

\item Given $(P_1, P_2)$,  the receiver $\RR_1$ computes $(\mu_1, \psi_2, \rho_2)$  and sends  $\mu_1$ and $(\psi_2 - \rho_2)$ to $\TT_1$ and $\TT_2$, respectively. Likewise, $\RR_2$ computes $(\mu_2, \psi_1,  \rho_1)$ and feeds back     $\mu_2$ and $(\psi_1 - \rho_1)$ to $\TT_2$ and $\TT_1$, respectively.

\item The transmitters $\TT_1$ and $\TT_2$ update $P_1$ and $P_2$, respectively, using \eqref{Eq:CapSlope} and the following equation
\begin{equation}
P_m(k+1)  = \min\l\{\l[P_m(k) + \frac{\partial A(P_1, P_2)}{\partial P_m}\Delta \gamma\r]^+, P_{\max}\r\}\nn
\end{equation}
where $k$ is the iteration index and $\Delta\gamma$ a  step size. 

\item Repeat Steps $2)-4)$ till the maximum number of iterations is performed or the changes on $(P_1, P_2)$ are sufficiently small. 
\end{enumerate}

\subsection{Outage Probability Criterion} \label{Section:OutageProb}
As the problem of minimizing $\Pout$ in \eqref{Eq:Pout} by power control is analytically intractable, the IPC algorithm is designed by minimizing an upper bound on $\Pout$. Using \eqref{Eq:Interf:Ub:c}, the SINR in \eqref{Eq:SINR} is lower bounded by $\widetilde{\SINR}_m^{[\ell]}$ where 
\begin{equation}\label{Eq:SINR:UB}
\widetilde{\SINR}_m^{[\ell]} :=  \frac{P_m \lambda_{mm}^{[\ell]}}{1 + P_nM\nu\lambda_{mn}^{[L-N+1]}\epsilon_n}, \quad m \neq n.
\end{equation}
Therefore, 
\begin{eqnarray}
\Pout &\leq&  \Pr\l( \min_{m=1,2} \widetilde{\SINR}_m^{[M]} < \theta \r). \label{Eq:Pout:LB}
\end{eqnarray}
Minimizing the above upper bound on $\Pout$ is similar in the mathematical structure to  the classic problem of optimal power control for single-antenna interference channels  \cite{Foschini:AutoPowerControl:1993}. 
The optimal transmission power for minimizing the right hand side of \eqref{Eq:Pout:LB} solves the following optimization problem 
\begin{equation}\label{Eq:PwrOptim}\begin{aligned}
&(P_1^\star, P_2^\star) = \arg\min_{P_1, P_2\in [0, P_{\max}]}(P_1, P_2) \\
&\textsf{s.t.}\ \min_{m=1,2} \widetilde{\SINR}_m^{[M]}(P_1, P_2) > \theta  
\end{aligned}
\end{equation}
where the first minimization implies that $(P_1^\star, P_2^\star)\preceq (P_1, P_2)$ for all $(P_1, P_2)$ that satisfies the constraint in \eqref{Eq:PwrOptim} as well as  $(0, 0) \preceq (P_1, P_2) \preceq (P_{\max}, P_{\max})$, called \emph{feasible} power pairs. 
Using \eqref{Eq:SINR:UB} and \eqref{Eq:PwrOptim}, the constraint in \eqref{Eq:PwrOptim} can be written as 
\begin{equation}\label{Eq:QoS}
P_m \geq a_m + b_{mn}P_n,\quad m\neq n 
\end{equation}
where $a_m := \frac{\theta}{\lambda_{mm}^{[M]}}$ and $ b_{mn} := \frac{M\nu\lambda_{mn}^{[L-N+1]}\epsilon_n\theta}{\lambda_{mm}^{[M]}}$. 
The minimum power  $(\acute{P}_1, \acute{P}_2)$ that satisfies the constraints in \eqref{Eq:QoS} is 
\begin{equation}\label{Eq:OptimPwr}
\acute{P}_m := \frac{a_m+b_{mn}a_n}{1-b_{mn}b_{nm}}, \quad m = 1,2. 
\end{equation}
This  expression gives  optimal power control as stated below.
\begin{proposition} \label{Prop:TxPwr} If $(P_1^\star, P_2^\star)$ in \eqref{Eq:PwrOptim} exists,  $(P_1^\star, P_2^\star) = (\acute{P}_1, \acute{P}_2)$ with $(\acute{P}_1, \acute{P}_2)$ given in  \eqref{Eq:OptimPwr}. 
\end{proposition}

Based on Proposition~\ref{Prop:TxPwr},  the corresponding   IPC feedback procedure is obtained  as follows. 

\noindent {\bf Algorithm $\mathbf{2}$:}

\begin{enumerate}
\item The receiver  $\RR_1$  computes $b_{12}$ and transmits $b_{12}$ to $\TT_2$. Similarly, $\RR_2$  computes $b_{21}$ and feeds back  $b_{21}$ to $\TT_1$. 

\item The receiver  $\RR_1$  computes $a_1$  and communicates $a_1$ to $\TT_1$. Likewise, $\RR_2$  computes $a_2$  and feeds back $a_2$ to $\TT_2$.  

\item The transmitters $\TT_1$ and $\TT_2$ compute $\acute{P}_1$ and $\acute{P}_2$, respectively and    set their transmission power equal to $(\acute{P}_1, \acute{P}_2)$ if they are feasible. Otherwise, arbitrary transmission power is used. 
\end{enumerate}

\section{Simulation Results}\label{Section:Simulation}
In the simulation, codebooks for quantizing $\bF^{\mathsf{i}}_1$ and $\bF^{\mathsf{i}}_2$ are randomly generated and   have equal sizes. The simulation parameters are set as  $L=6$, $M=2$, $N=3$, and $B=6$. The interference margin is either fixed at $\tau = 2$ or increased  as $\tau = 0.4P_{\max}$.

\begin{figure}
\centering
\vspace{-10pt}\includegraphics[width=8.5cm]{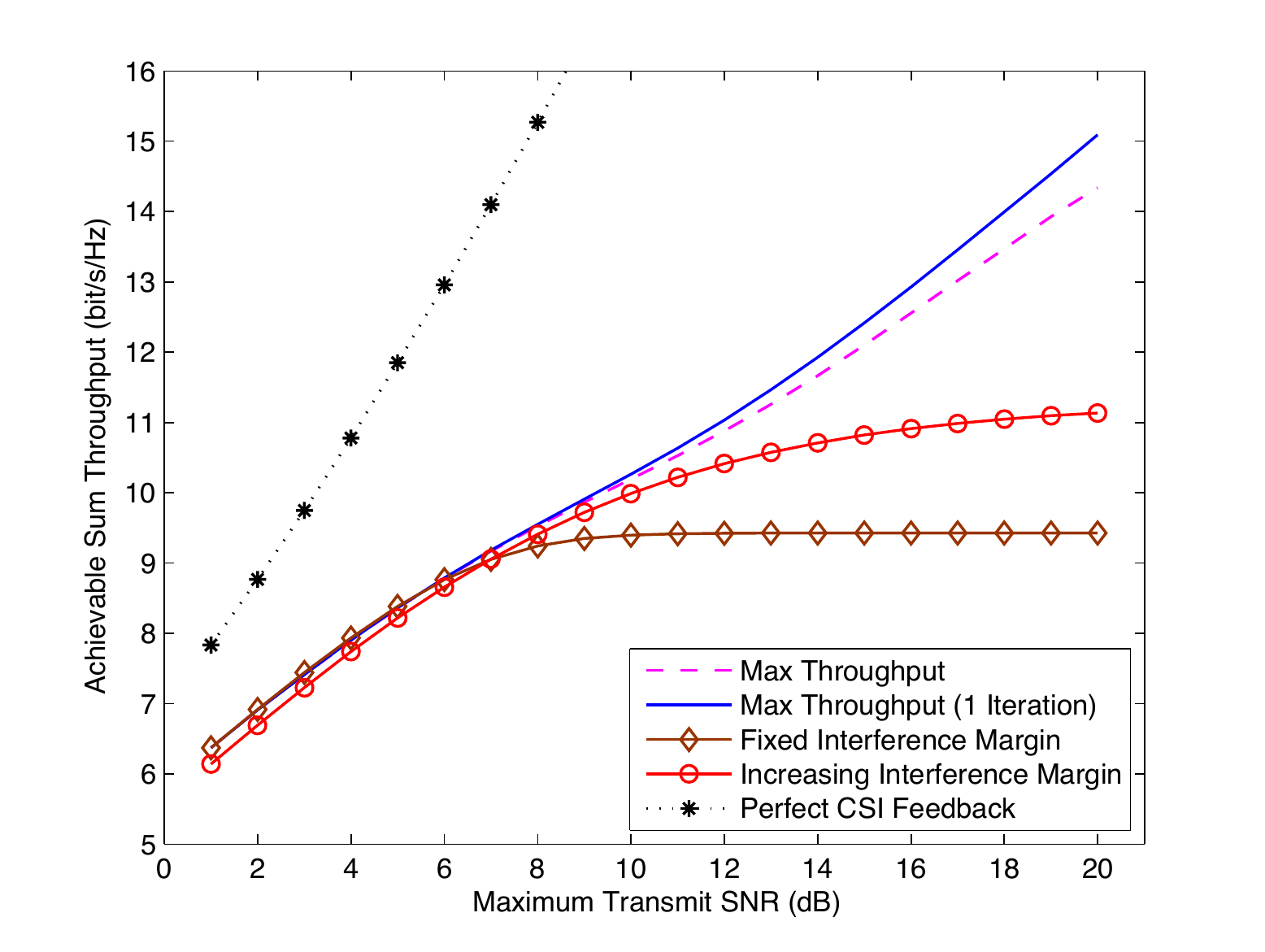}\vspace{-12pt}
  \caption{Comparison of achievable sum throughput between different IPC feedback algorithms for the coupling factor $\nu=0.2$.  }\label{Fig:IPC:SumCap}
\end{figure}

Fig.~\ref{Fig:IPC:SumCap} compares the achievable throughput of different IPC feedback algorithms. Significant   coupling ($\nu = 0.2$) between links  is observed to decrease achievable throughput dramatically with respect to perfect CSI feedback. For large $P_{\max}$, the  IPC feedback Algorithm~$1$ designed for maximizing the achievable throughput is observed to provide substantial  throughput gain over those based on interference margins.  Furthermore, increasing $\tau$ with growing $P_{\max}$ gives higher throughput than fixed $\tau$, which is consistent with Lemma~\ref{Lemma:SumCap:IM}. 

\begin{figure}
\centering
\vspace{-10pt}\subfigure{\includegraphics[width=8.5cm]{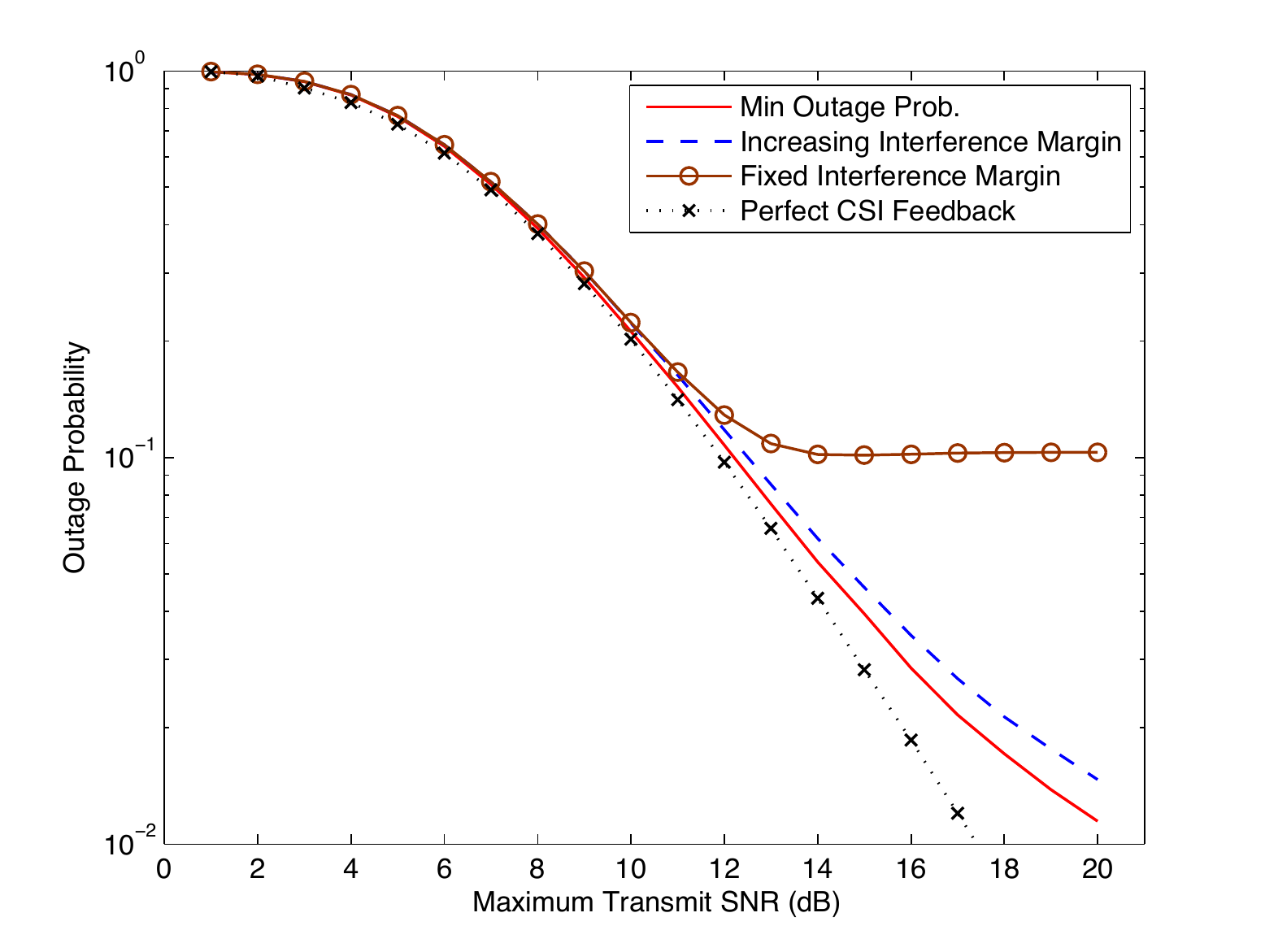}}\vspace{-15pt}\\
\subfigure{\includegraphics[width=8.5cm]{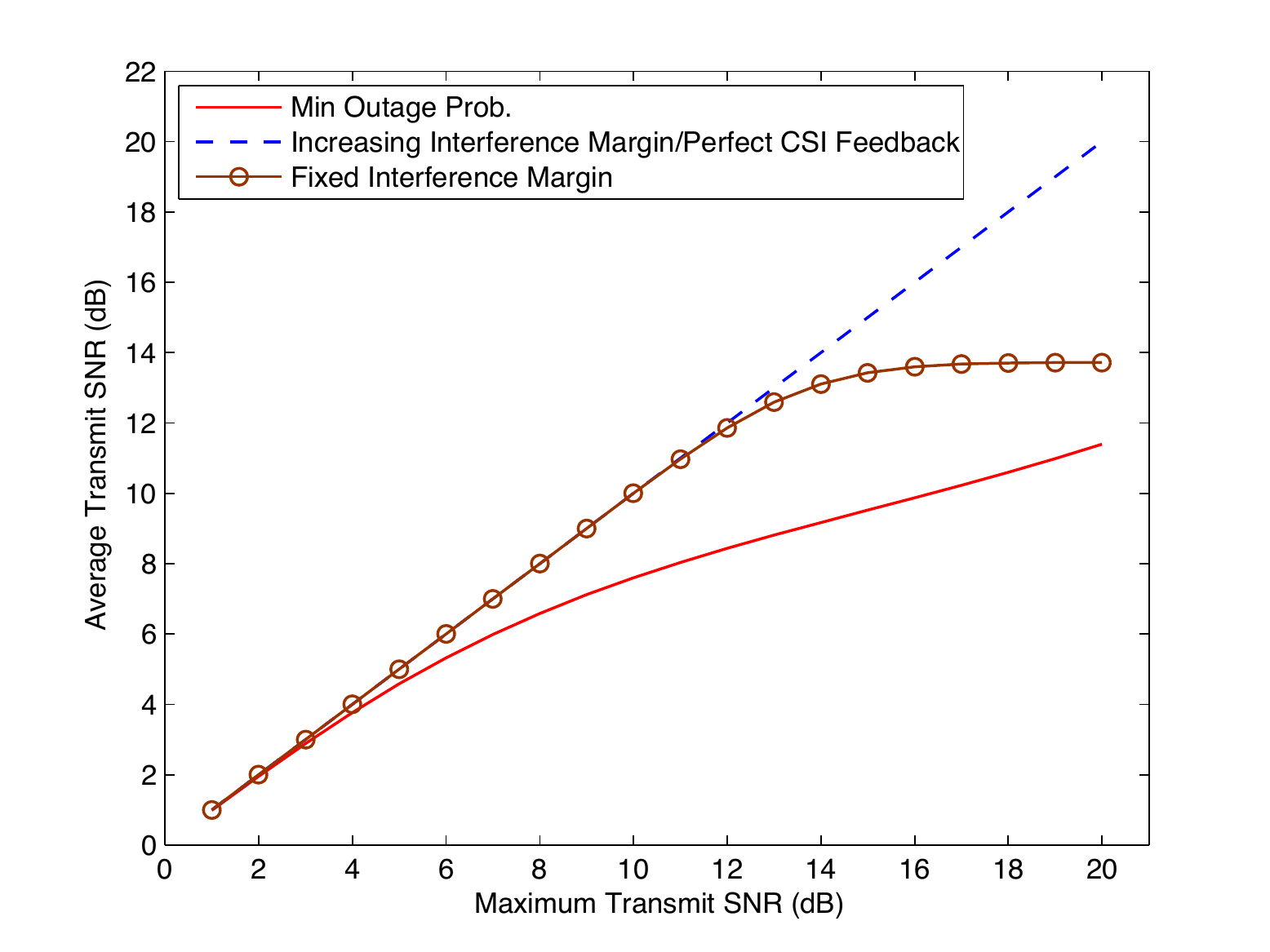}}\vspace{-12pt}
  \caption{Comparison of outage probability (upper) and average transmit SNR (lower) between different IPC feedback algorithms for the coupling factor $\nu = 0.05$.   }\label{Fig:IPC:Outage}
\end{figure}

Fig.~\ref{Fig:IPC:Outage} compares the outage probabilities and average transmit SNRs  of different IPC feedback algorithms. With respect to the IPC feedback using increasing $\tau$  or with perfect feedback, Algorithm~$2$   dramatically decreases average transmission power. Moreover,  Algorithm~$2$ yields lower  outage probability than the two algorithms using $\tau$. 
Finally, fixing $\tau$ causes $\Pout$ to saturate as $P_{\max}$ increases.  

\bibliographystyle{ieeetr}

\end{document}